\documentclass[pra, 10pt, twocolumn, tightenlines, longbibliography, aps,superscriptaddress,nofootinbib]{revtex4-2}
\usepackage{amsmath}
\usepackage{graphicx}
\usepackage{color}
\usepackage{amssymb}
\usepackage[utf8]{inputenc}
\usepackage[T1]{fontenc}

\usepackage{comment}
\usepackage[normalem]{ulem}
\usepackage{hyperref}
\usepackage{url}
\usepackage{cleveref} 
\hypersetup{colorlinks,
 linkcolor=blue,%
 citecolor=blue,%
 urlcolor=blue
}

\usepackage[normalem]{ulem} 

\definecolor{teal}{rgb}{0,0.5,0.5}

\definecolor{capri}{rgb}{0,0.75,1}

\usepackage{float}
\crefname{figure}{Fig.}{Figs.}
\begin{document}

\title{Pattern formation in ring condensates subjected to bichromatic driving}

\author{Premabrata Manna}
\affiliation{Department of Physics, Indian Institute of Technology Guwahati, Guwahati 781039, Assam, India}

\author{S. I. Mistakidis}
\affiliation{Department of Physics and LAMOR, Missouri University of Science and Technology, Rolla, MO 65409, USA}

\author{P. G. Kevrekidis}
\affiliation{Department of Mathematics and Statistics, University of Massachusetts Amherst, Amherst, Massachusetts 01003-4515, USA}

\affiliation{Department of Physics, University of Massachusetts Amherst, Amherst, 01003, MA, USA}

\affiliation{Theoretical Sciences Visiting Program, Okinawa Institute of Science and Technology Graduate University, Onna, 904-0495, Japan}

\author{Pankaj Kumar Mishra}
\affiliation{Department of Physics, Indian Institute of Technology Guwahati, Guwahati 781039, Assam, India}

\date{\today}

\begin{abstract} 
We investigate the dynamical formation of nonlinear patterns in one-dimensional ring condensates under bichromatic periodic modulation of the interaction strength. 
The stability phase diagram of the condensate's  homogeneous density state is analytically derived through a suitable biharmonic variant of the Mathieu equation and computing the associated Floquet spectrum. 
It reveals the complex interplay 
between the driving parameters, i.e.,  amplitude, frequencies, and the so-called frequencies' mixing angle, which dictate the instability onset and the selective  enhancement of higher-order resonance tongues, thus offering precise control over the excited modes. 
These results are in agreement with time-dependent mean-field simulations evidencing the emergence of density wave modulations of specific momenta, while enabling a deeper understanding of the nonlinear stage of the relevant
instability.
Further insights on the ensuing unstable nonlinear dynamics are provided through a reduced {five-mode} model which captures the instability onset, the  oscillatory behavior of the mode populations and the phase-space dynamics, in agreement with the mean-field predictions. 
Our study highlights the versatility of bichromatic driving to generate and control  complex nonlinear patterns that are 
within reach in present day ultracold atom  experiments. 
\end{abstract}

\maketitle

\section{Introduction}\label{sec:intro}

The spontaneous emergence of patterns in driven systems is a universal phenomenon observed across many fields, including hydrodynamics, solid-state physics, nonlinear optics, chemistry, and biology~\cite{Cross1993,Cross2009}. 
A paradigmatic example is the formation of Faraday waves -- standing surface patterns that appear in a vertically shaken fluid -- originally reported by Michael Faraday~\cite{Faraday1831}. 
These waves emanate from parametric instabilities and provide a prototypical framework for studying structure formation via spontaneous symmetry breaking. 
Similar phenomena have been observed in diverse systems, ranging from classical fluids~\cite{Ciliberto1984,Binks1997,adou2016faraday, li2018linear, maity2020instability, kumar1996linear}, to multimode lasers~\cite{Szwaj1998}, and superfluid helium~\cite{Abe2007}.

Ultracold atomic gases, and in particular Bose–Einstein condensates (BECs), have emerged as a powerful platform for exploring such driven quantum phenomena in a highly controllable environment~\cite{bloch2008many, kevrekidis2004pattern}. 
Parametric driving in BECs can be implemented through various techniques, including temporal modulation of the trapping potential~\cite{GarcaRipoll1999,Staliunas2004,Nicolin2007,Balaz2012} or by tuning the interatomic interactions periodically via Feshbach resonances~\cite{Inouye1998,Staliunas2002,FRM}. 
These techniques have enabled the experimental observation of Faraday patterns in quasi-one-dimensional elongated single-component condensates~\cite{Engels2007} and two-dimensional ones~\cite{kwon2021spontaneous}. 
In two-dimensions Bose fireworks~\cite{clark2017collective,fu2018density}, and the dynamical formation of supersolid-like configurations~\cite{liebster2025supersolid} have also been seen to emerge under monochromatic interaction driving.

Beyond scalar BECs, theoretical studies have extended the concept of Faraday pattern formation to more complex systems, such as two-component BECs~\cite{Chen2019,Balaz2012,Zhang2022}, spin-1 BECs~\cite{kargudri2025faraday}, dipolar BECs~\cite{Nath2010,Lakomy2012,Nadiger2024,Vudragovi2019,Turmanov2020}, and Bose-Fermi  mixtures~\cite{Abdullaev2013}. 
While these previous studies focused on harmonic traps and box-potentials, more recently ring-shaped condensates have been shown to provide an alternative geometry for exploring non-linear pattern formation~\cite{zhu2019,haberichter}, whose experimental exploration
remains an exciting frontier~\cite{campbell}.

Naturally, an additional layer of complexity arises when systems are driven at two distinct competing  frequencies. 
In classical fluids, two-frequency forcing can stabilize or destabilize patterns and give rise to novel structures, including square and hexagonal lattices~\cite{Edwards1994,Besson1996,Zhang1997,Silber1999,Arbell2002,Topaz2004,Takagi2015}.  
In quantum gases, however, the impact of bichromatic driving has been thus far much less explored. In this context, recent experimental work in two-dimensional Cs condensates revealed the spontaneous nucleation of complex matter-wave symmetric lattice structures~\cite{Zhang2020}. 

As such, the potential of bichromatic driving to controllably generate patterns 
constitutes a topic worth exploring, as a complement towards
such recent experiments and motivation to further ones
along this vein.

It is this critical gap that our work aims to address. 
More specifically, we explore the emergence of pattern formation in a one-dimensional (1D) BEC trapped in a ring geometry and being exposed to a bichromatic driving protocol of the interaction strength. 
Deploying Floquet theory, we analytically derive the associated generalized Mathieu equation and construct the underlying stability phase diagram of homogeneous density states identifying resonance tongues and their dependence on the driving parameters. 
It is shown that the ratio of the two driving frequencies and 
the parameter that controls the relative strength of their 
contribution 
(that we refer to as the mixing angle--- see details below) can be used 
for tailoring the ensuing parametric instabilities. 
In particular, we explicate that tuning these parameters can selectively enhance higher-order resonance tongues while narrowing the primary instability band, enabling controlled excitation of different modes. Indeed, we propose a protocol that
homotopically transitions from the parametric resonances
associated with the first frequency to those associated
with the second (or vice versa).
These analytical predictions are validated by time-dependent Gross–Pitaevskii simulations, which uncover the dynamical formation of spatially periodic density patterns of discrete wavenumbers imposed by the ring geometry. Interestingly, these simulations
reflect also the nonlinear stage (and eventual 
saturation, as well as oscillatory evolution) of the
corresponding instability.
Finally, we introduce a reduced {five-mode} model that adequately reproduces the key nonlinear features of the condensate mean-field dynamics, including the onset of the instability, the mode population oscillations and the corresponding phase-space trajectories.

Our work unfolds as follows. In~\cref{sec:MF}, we present the mean-field partial differential equation model corresponding to the condensate confined in the ring geometry. In~\cref{sec:mathieu}, using Floquet theory on the Mathieu equation we analytically extract the stability phase diagram delineating the parametric regions of instability in the presence of bichromatic driving in the interaction strength. Section~\ref{sec:numerical} discusses the properties of the instability through mean-field Gross-Pitaevskii simulations, i.e., it presents our field-theoretic numerical
results. 
Section~\ref{sec:five_Mode} elaborates on the construction of a reduced 
{five-mode} model to predict the instability characteristics and provides comparisons with the respective mean-field predictions. We conclude and discuss future perspectives based on our work in  Sec.~\ref{sec:con}. 
Appendix~\ref{A:1} elaborates on the  Floquet analysis used on the generalized  Mathieu equation to investigate the system’s stability properties.

\section{Mean-field model and driving protocol}\label{sec:MF}

\subsection{Setup and ring confinement}

We consider a BEC confined in a 1D toroidal trap of radius $R$. 
It is strongly confined along the radial and axial ($r$ and $z$) direction and remains effectively uniform along the toroidal ($\theta$-) direction. Toroidal traps can be experimentally designed by using a blue-detuned laser as a “plug beam” in a magnetic trap with a harmonic potential  \cite{Ryu2007, Qin2019, Halkyard:2010hoa}. 
Adjustment of the plug beam’s waist and intensity yields variations on the trap's radius and other confinement parameters.

The corresponding 3D toroidal trapping potential is given by 
\begin{equation}
V(r, \theta, z) = \frac{1}{2} M \Omega_z^2 z^2 + \frac{1}{2} M \Omega_a^2 r^2 + V_0 \exp \left( -\frac{2r^2}{w_0^2} \right),
\end{equation}
where $M$ is the mass of the atomic species of interest, $\Omega_{x}=\Omega_{y} \equiv \Omega_{a}$, and $\Omega_{z}$ refer to the angular trapping frequencies along the $x-$, $y-$, and $z-$ directions. 
The waist of the beam is $w_{0}$, 
while $V_0$ is related to the power of the plug-beam, whose minimum is located at \( z = 0 \) and \( r = R \), 
where the radius reads
\begin{equation}
R^2 = \frac{w_0^2}{2} \ln \left( \frac{4V_0}{M \Omega_a^2 w_0^2} \right).
\end{equation}

Taylor expanding the trapping potential around  $r = R$, and keeping terms up to quadratic order (i.e., neglecting terms of the order of $(r - R)^3$ and higher), the potential can be re-written as
\begin{equation}
V(r, \theta, z) = \frac{1}{2} M \Omega_z^2 z^2 + \frac{1}{2} M \Omega_r^2 (r - R)^2,
\end{equation}
where \( \Omega_r = 2\Omega_a \, R/w_0 \) 
represents the effective radial trap frequency. 
Under strong radial and axial confinement, the condensate dynamics becomes effectively 1D along the angular coordinate $\theta$.  
Nevertheless, it is relevant to mention that the considerations
provided below are directly applicable also to a genuinely
1D setting, as concerns the relevant phenomenology.

\subsection{Effective Hamiltonian}

In the limit of \(\sqrt{\hbar/M\Omega_r}\), \(\sqrt{\hbar/M\Omega_z} \ll R\), the dimensionless Hamiltonian for repulsive interactions can be expressed as~\cite{Halkyard:2010hoa,zhu2019} 

\begin{align}
\hat{H} = \int_{0}^{2\pi} d\theta \,\Bigg[
& -\lambda\, \hat{\psi}^{\dagger}(\theta) \frac{\partial^2}{\partial \theta^2} \hat{\psi}(\theta) \notag \\
& + \frac{\pi g(t)}{N} \hat{\psi}^{\dagger}(\theta) \hat{\psi}^{\dagger}(\theta) \hat{\psi}(\theta)
\Bigg],
\label{eq:hamiltonian}
\end{align}
where $\hat{\psi}$ is the condensate's field operator, $\theta$ denotes the azimuthal coordinate, and the parameter $\lambda = \pi \hbar / (2MNR\sqrt{\Omega_r \Omega_z} a_s)$. The effective 1D interaction strength is $\tilde{g} = 2N\hbar \sqrt{\Omega_r \Omega_z} a_s/R$, with $a_s$ being the 3D scattering length and $N$ is the total atom number. In the Hamiltonian of Eq.~(\ref{eq:hamiltonian}), we take $R$ as the unit of length, while energy and time are expressed in terms of \(\epsilon_0 = |\tilde{g}|/2\pi\) and \(t_0 = 2\pi \hbar/|\tilde{g}|\) respectively. 

In a corresponding experiment, our setup can be realized, for instance, by a \(^{87}\mathrm{Rb}\) condensate with $N=2600$ atoms being the particle number considered throughout our analysis below.  
The background scattering length can be around \(a_{s}=98.09\,a_{0}\), while the  
radial (axial) trap frequency \(\Omega_{r}=2\pi\times540~\mathrm{Hz}\) [\(\Omega_{z}=2\pi\times338~\mathrm{Hz}\)]  
and the ring radius \(R=10~\mu\mathrm{m}\). 
Under these conditions, the dimensionless parameter  
\(\lambda=2\pi/2000\), and the corresponding healing length corresponds to 
$\xi=\left( \frac{4 M N \sqrt{\omega_{r}\Omega_{z}} a_{s}}
{\pi R \hbar}\right)^{-1/2}=0.396~\mu\mathrm{m}$. 

\begin{figure*}[!htp]
\includegraphics[width=\textwidth]{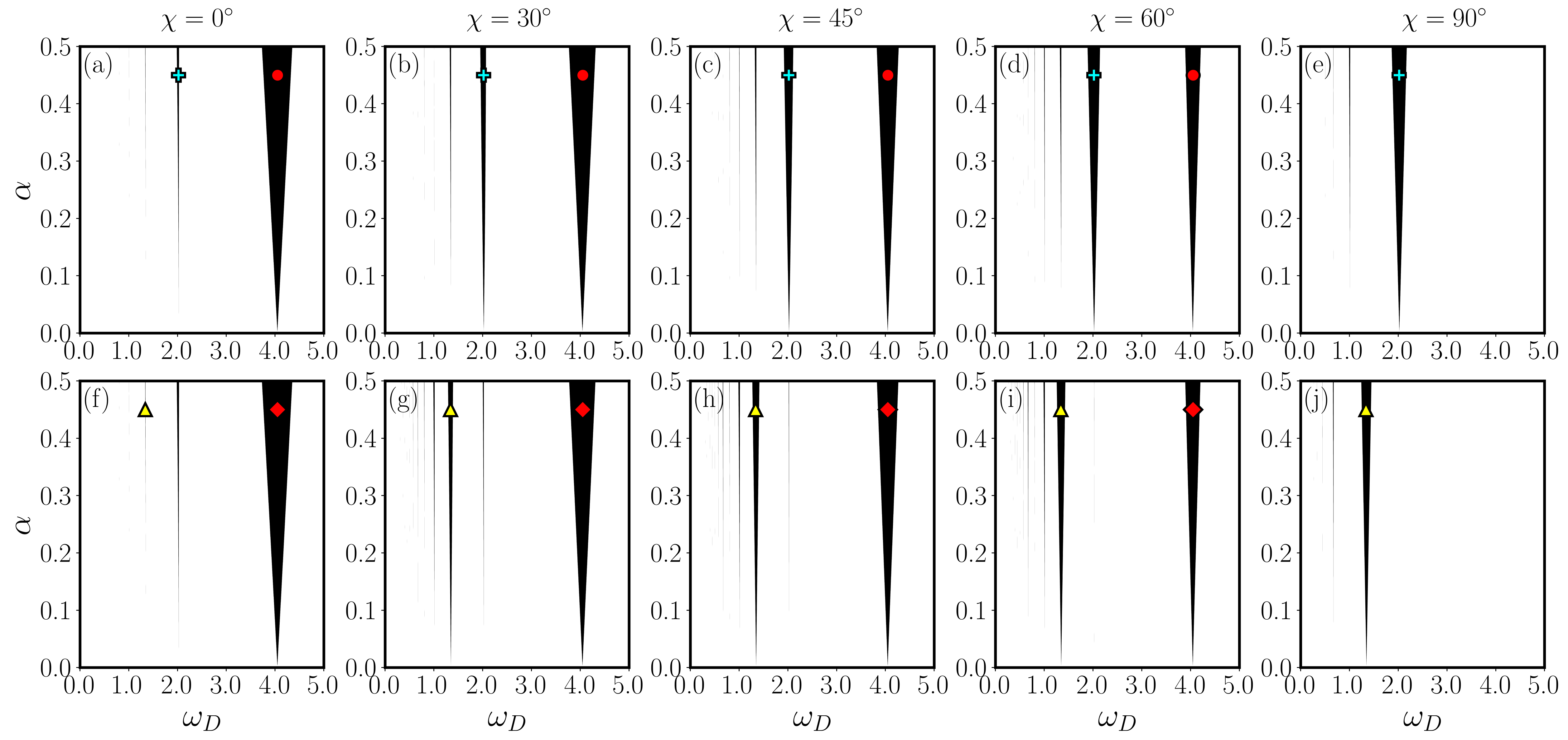} 
\caption{Effect of frequency ratio and mixing angle
$\chi$ [in Eq.~\eqref{eq:interaction_modulation}] 
on the stability phase diagram.  
Floquet resonance tongues (dark shaded regions) across the $\alpha$-$\omega_D$ plane for a 1D BEC under bichromatic interaction driving for unstable wavenumber $ k = 20 $ and different frequency ratios and mixing angles, $\chi$. 
Upper (lower) panels refer to frequency ratios $1{:}2$ ($1{:}3$), while each column from left to right corresponds to mixing angle $\chi = 0^\circ, 30^\circ, 45^\circ, 60^\circ, ~\textrm{and}~90^\circ $, respectively.   
For the $1{:}2$ frequency ratio and $\chi=0^\circ$, the first resonance tongue appears at $\omega_D = 2\Lambda(k)$ (red circles) and the second at $\omega_D = \Lambda(k) $ (cyan plus markers).
At $\chi=90^\circ$, the $m=2$ scenario emerges with the dominant resonance
now lying at  $\omega_D = \Lambda(k) $ (cyan plus markers). 
For the $1{:}3$ ratios and $\chi=0^\circ$, the first resonance tongue takes place at $\omega_D = 2\Lambda(k)$ (red diamonds)  and the third one 
occurs around $\omega_D = 2 \Lambda(k)/3 $ (yellow triangular marker).
Again at $\chi=90^\circ$, the $m=3$ setting leads to a dominant peak
at $\omega_D = 2 \Lambda(k)/3 $ (yellow triangular marker). For intermediate
angles, both for $m=2$ and for $m=3$, one sees a combination of the relevant
frequencies, progressively transitioning from those of $\chi=0^\circ$ to those of $\chi=90^\circ$.}
\label{fig:mathieu_double}
\end{figure*}

\subsection{Bichromatic driving protocol}

To induce parametric resonances in our system, we employ a bichromatic periodic modulation of the interaction strength following 
\begin{align}
g(t) = 1 +\alpha\left[\cos(\chi)\cos(\omega_D t)+\sin(\chi)\cos(m\omega_D t)\right].  
\label{eq:interaction_modulation}
\end{align}
Importantly, even with one of these frequencies alone,
one can observe parametric resonances, in line with 
earlier works~\cite{Staliunas2002} (see also~\cite{GarcaRipoll1999,Kevrekidis2000ParametricQuantumResonances}).
Here, $m$ is an integer and we define the mixing angle 
$\chi$ within the driving protocol of Eq.~\eqref{eq:interaction_modulation}. Indeed, in introducing the relevant angle, we 
are borrowing the relevant terminology from the fluid 
dynamics community~\cite{Besson1996}, where such biharmonic 
drives have been an especially popular topic~\cite{Edwards1994,Silber1999,Arbell2002}.
Throughout, we assume small driving amplitudes, i.e., $\alpha \ll 1$ and refer to the driving frequency ratio used in Eq.~(\ref{eq:interaction_modulation}) as the ratio $1{:}m$. The latter describes the fraction of the two involved driving  frequencies, $\omega_1=\omega_D$ and $\omega_2=m\omega_D$, dictating the interaction modulation. Experimentally, such a protocol can be realized by applying a sinusoidally oscillating magnetic field, which modulates the 3D scattering length connected with the 1D interaction strength ($g(t)$) of the atoms.  
A bichromatic scattering length modulation was, for instance, demonstrated in the experiment of Ref.~\cite{Zhang2020} and
hence is very much within the realm of what is experimentally
accessible.

\subsection{Gross-Pitaevskii equation}

At the mean-field level, the condensate driven dynamics is described by the dimensionless Gross–Pitaevskii equation (GPE) given by 
\begin{align}
i\frac{\partial}{\partial t} \psi(\theta,t) = \left[ -\lambda \frac{\partial^2}{\partial \theta^2} + 2\pi g(t) |\psi|^2 \right] \psi(\theta,t),
\label{eq:gpe}
\end{align}
where $\psi = \langle \hat{\psi} \rangle/\sqrt{N}$ is the condensate wave function being normalized as  $\int_{0}^{2\pi} |\psi|^2 d\theta = 1$. 
We remark that in the absence of interaction modulation, the condensate's ground state is uniform and stable for $\lambda<-2$ and $\lambda>0$. 
However, when $-2<\lambda<0$, the ground state supports a soliton solution~\cite{Kanamoto2003}. In this work, we focus on the parameter regime $\lambda>0$, i.e., involving positive
scattering lengths and repulsive interactions (self-defocusing
nonlinearities).


\section{Linear stability analysis with double frequency modulation}\label{sec:mathieu}

We now analyze the parametric resonance of a BEC confined in a 1D ring trap, when the interaction strength is periodically modulated with two driving frequencies, as described above. 
In direct analogy to classical fluids, temporal modulation can trigger instabilities when the driving frequency lies within specific resonance bands~\cite{maity2020parametrically}, as
per the well-known theory of parametric resonances within
Hill's equations~\cite{magnus2004hill}. 
These unstable regions are characterized by the so-called instability tongues~\cite{Rand2018} (sometimes referred
to as Arnold or Floquet tongues), which describe the parametric regions where small perturbations grow exponentially. 
Our goal here is to derive the effective (generalized)
Mathieu equation for the condensate under the biharmonic
drive, determine the resonance conditions, and explain how the 
two-frequency setting modifies the stability structure as compared to the single frequency driving and how it can be leveraged to control the instability regimes. 
The spatially homogeneous, temporally periodic solution of Eq.~(\ref{eq:gpe}) is given by $\psi_{0}\left(t\right)=\frac{1}{\sqrt{2\pi}}\exp[-i\theta_p\left(t\right)]$, where 
$\theta_p = \int_0^t g(s) ds$ denotes the phase of the condensate wavefunction. 

To examine its stability against spatially modulated perturbations, i.e., whether 
infinitesimal spatial modulations will exponentially grow
or stay bounded, we undertake  linear stability analysis. 
For this reason, as described in Refs.~\cite{Staliunas2002,Nicolin2007,PhysRevA.108.063317}, we introduce a small spatial perturbation 
\begin{equation}
\psi(\theta , t) = \psi_{0}\left(t\right)[1 + (u(t)+iv(t))\cos(k\theta)], 
\end{equation}
where $u(t)$ and $v(t)$ represent the real and imaginary parts of the perturbation amplitude, while $k$ is the perturbation wavenumber. 

Substituting this ansatz into Eq.~(\ref{eq:gpe}) and linearizing in terms of the perturbation amplitudes ($u(t)$, $v(t)$), 
(and then combining the two equations for $u$ and $v$ into
a single equation for $u(t)$)
one obtains the following generalized Mathieu equation for $u(t)$ given by 
\begin{align}
\frac{d^2 u}{dt^2} + \Big[ \Lambda^2(k) + &2\alpha\lambda k^2 \big( \cos(\chi)\cos(\omega_D t) \notag \\
& +\sin(\chi)\cos(m\omega_D t)\big) \Big] u = 0. 
\label{eq:mathieu_equation}
\end{align}

In this expression, \(\Lambda(k) = k\sqrt{\lambda^2k^2 + 2\lambda}\) is the natural frequency in the absence of scattering length modulation (i.e., $\alpha=0$). 
The above generalized Mathieu equation represents a parametrically driven oscillator with a natural frequency $\Lambda(k)$ featuring a series of resonance tongues.
When a single drive is present in the case of the standard
Mathieu equation, say, for $\chi=0$, it is well-known~\cite{Rand2018} that the resonances occur
at {frequencies}
\begin{align}
{\Lambda(k)}=n \big(\frac{\omega_D}{2}\big)  \quad \text{with } n = 1, 2, 3,\ldots
\label{eq:omega_n}
\end{align}
where $n=1,2, \dots$ designates the location of the first, second,$\dots$ resonance (Mathieu/instability) tongue. Since $m$ is an integer Eq.~(\ref{eq:mathieu_equation}) admits periodic solutions of period $T=2\pi/\omega_D$. These can be determined through the lens of Floquet theory which will  subsequently allow to analyze the system’s stability~\cite{Rand2018,barone1977floquet}, see also Appendix~\ref{A:1} for derivation details. 
Moreover, a careful inspection of Eq.~(\ref{eq:mathieu_equation}) unveils that the onset of the instability depends on driving amplitude $\alpha$, frequency $\omega_D$, and unstable wavenumber $k$ which is an integer due to the considered ring geometry~\cite{zhu2019}. Hence, variations of these driving parameters may facilitate characterization and control of the ensuing stability behavior of the original homogeneous BEC state.


We will now consider a systematic variation of the
mixing angle $\chi$ and explore how the nature of the
resonances changes as $\chi$ changes.
Applying a single frequency driving, namely in our case $\chi=0^\circ$,  Eq.~(\ref{eq:mathieu_equation}) reduces to the standard Mathieu form~\cite{Rand2018,Staliunas2002} which reads 
\begin{align}
\frac{d^2 u}{dt^2} + \Big[ \Lambda^2(k) + 2\alpha\lambda k^2 \cos(\omega_D t)  \Big] u = 0. 
\label{eq:mathieu_equation_w}
\end{align} 

Figure~\ref{fig:mathieu_double}(a) shows the stability phase diagram for single frequency driving (i.e., $\chi=0^\circ$) in the parametric plane $\alpha$-$\omega_D$ for unstable wavenumber fixed to $k=20$. 
Note that here if any other value of the wavenumber $k$ is chosen, it yields the same qualitative stability features that will be described below. 
Parametric resonances at a discrete $k$ (enforced by the ring geometry) occur  as long as $\omega_D$ lies inside the unstable regions predicted by the Mathieu equation [Eq.~\ref{eq:mathieu_equation_w}]. 
Specifically, the stability of the condensate is analyzed by computing the monodromy matrix over one driving period and examining its eigenvalues, known as Floquet multipliers~\cite{bukov2015universal,magnus2004hill}. If the eigenvalues are larger than unity, then instability occurs. Further details related to this method are provided in Appendix~\ref{A:1}. 
In this single frequency driving case,  the primary resonance tongue ($n=1$) clearly dominates, while higher-order tongues appear to be much narrower and typically suppressed in accordance to previous findings on driven condensates~\cite{nguyen2019parametric,Engels2007}.

\begin{figure}[t] 
 
\includegraphics[width=\linewidth]{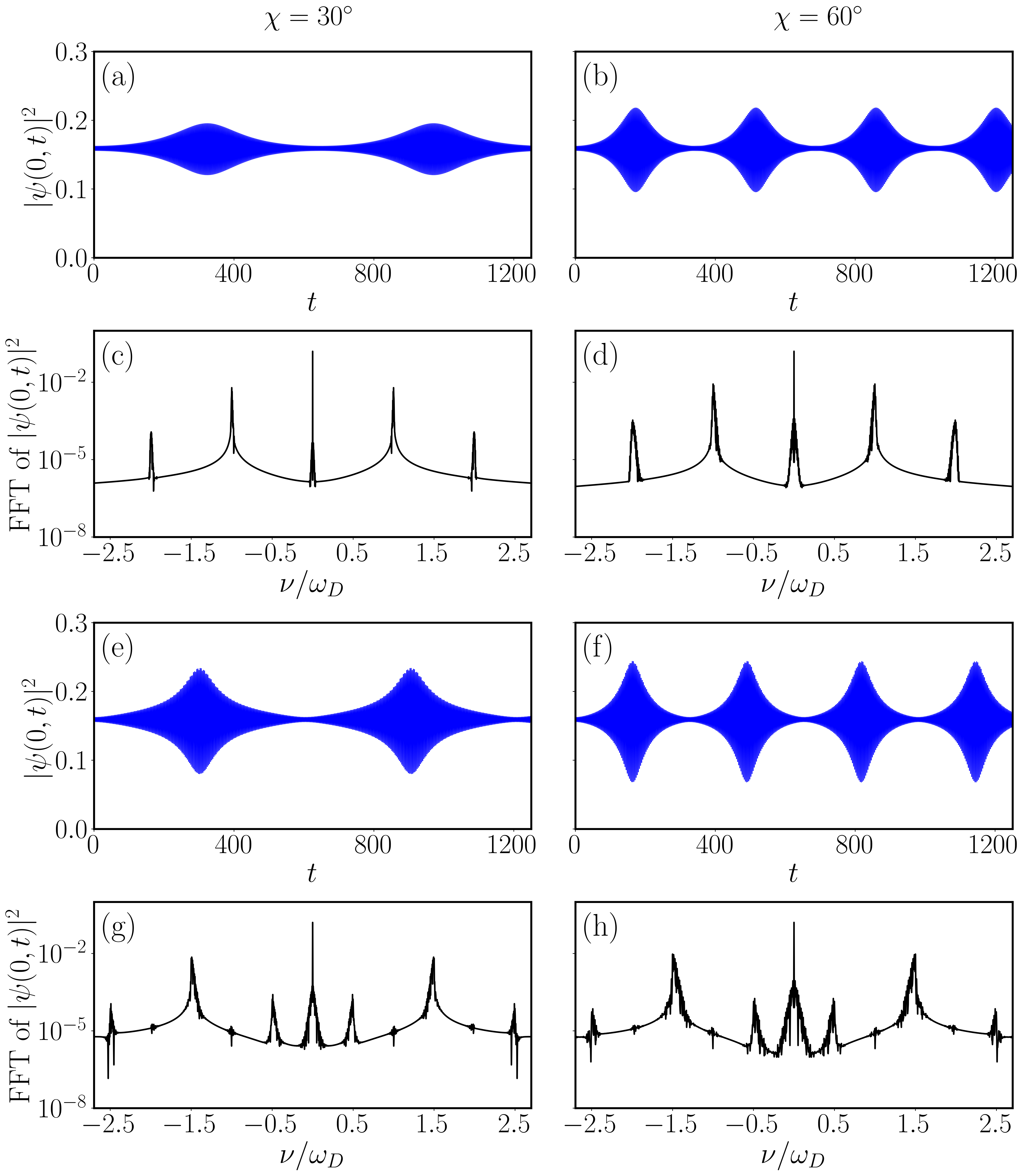} 
\caption{Time-evolution of the driven condensate density at $\theta=0$, $|\psi(0,t)|^2$, and its Fourier spectra within different resonance tongues for $k=20$. 
Panels (a)–(b) show the dynamics for the second resonance tongue ($\omega_D = 2.03$) under a $1{:}2$ double-frequency drive with $\chi = 30^\circ$ and $\chi = 60^\circ$, respectively. 
The corresponding Fourier spectra are shown in (c)–(d). 
Dynamics of $|\psi(0,t)|^2$ within the third resonance tongue ($\omega_D = 1.34$) under a $1{:}3$ bichromatic driving  for (e) $\chi = 30^\circ$ and (f) $\chi = 60^\circ$. 
The respective Fourier spectra of panel (e) [(f)] are displayed in (g) [(h)].  
Increasing $\chi$ with fixed frequency ratio enhances the instability amplitude, see panels (a) and (b). 
The temporal period of the patterns is modified by tuning the frequency ratio, e.g. compare panels (c) and (g). Here $\nu/\omega_D$ is the Fourier frequency $\nu$ divided by the external driving frequency $\omega_D$.
In all cases, the driving amplitude is fixed at $\alpha = 0.09$.}
\label{fig:2_omega_1}
\end{figure}

The stability of the system, dictated by Eq.~(\ref{eq:mathieu_equation}), changes {\it dramatically} under bichromatic driving.  
Specifically, the resonance tongues captured by Floquet theory in the presence of bichromatic modulation of the interactions and for the wavenumber $k=20$  are presented in \cref{fig:mathieu_double} for frequency ratios $1{:}2$ (top row) and $1{:}3$ (bottom row). 
Each column corresponds to different mixing angles,  namely $\chi = 0^\circ, 30^\circ, 45^\circ, 60^\circ,$ and $90^\circ$. 
In all cases, the instability windows (see shaded areas) have the characteristic triangular tongue shape with their apexes being determined by the parametric resonance condition of Eq.~(\ref{eq:omega_n}) for $\chi=0$ . 

This behavior can be clearly seen in 
Fig.~\ref{fig:mathieu_double}(a), (d), 
where, for instance, 
the first tongue appears near $\omega_D \approx 2\Lambda(k)$ ($\omega_D \approx 4.05$), the second at $\omega_D \approx \Lambda(k)$ ($\omega_D \approx 2.02$), and   
the third occurs (very faintly) around $\omega_D \approx 2\Lambda(k)/3$ ($\omega_D \approx 1.35$). {For the \(1{:}2\) and \(1{:}3\) frequency-driving cases, representative frequencies are selected as \(\omega_D = 2.03\) (second Mathieu tongue) and \(\omega_D = 1.34\) (third Mathieu tongue), respectively, for the analysis presented in the following sections.
}

\begin{figure*}[t]
\includegraphics[width=\textwidth]{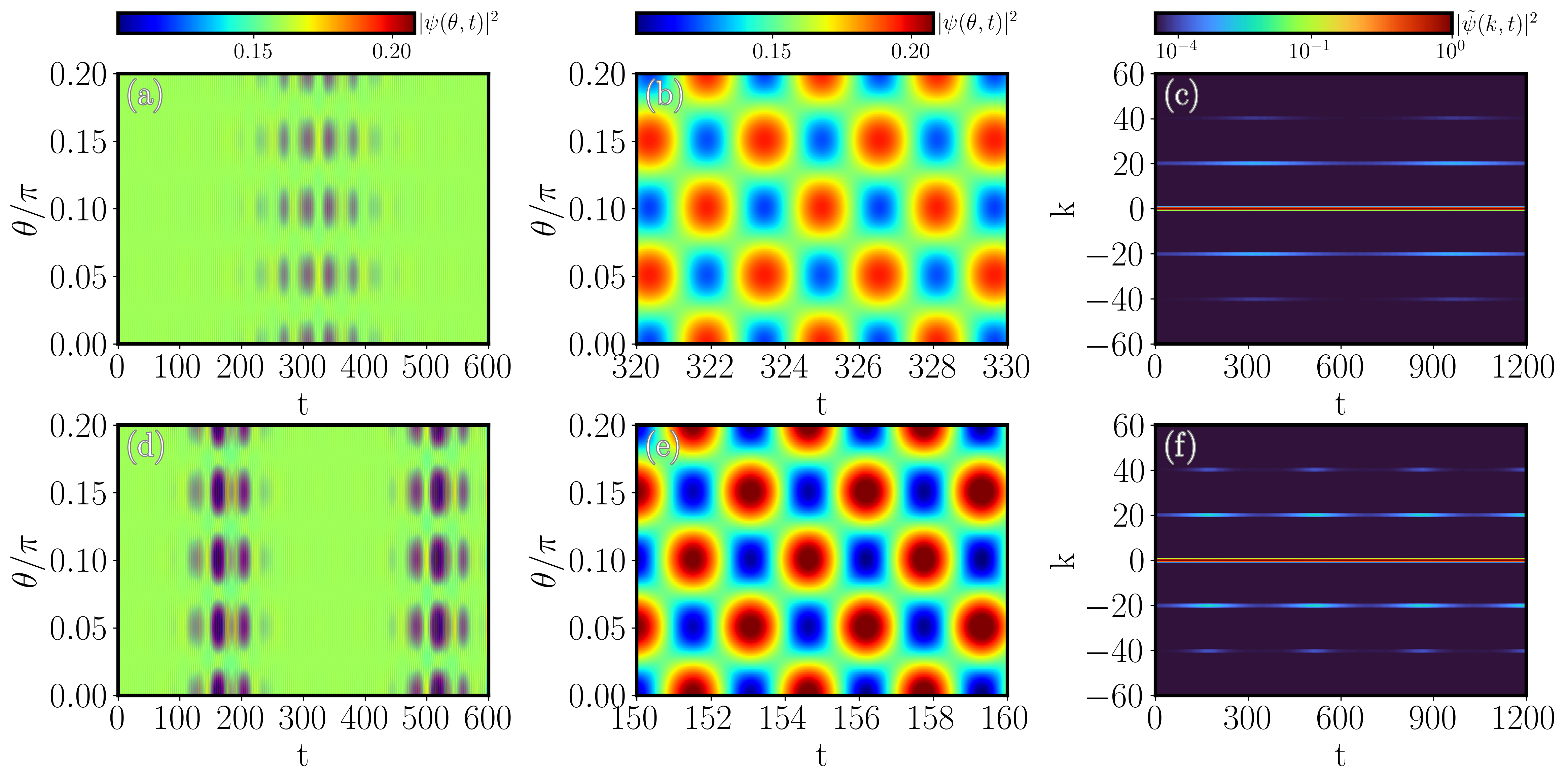} 
\caption{(a)-(e) Spatio-temporal density evolution within the GPE following bichromatic driving of the interaction strength [Eq.~(\ref{eq:interaction_modulation})] with characteristics $\omega_D=2.03$ (second tongue), frequency ratio $1{:}2$, amplitude $\alpha = 0.09$, and wavenumber $k = 20$.   
Panels correspond (a)–(c) to mixing angles $\chi = 30^\circ$, and (d)–(f) to   $\chi = 30^\circ$
(b) [(e)] Density evolution shown in (a) [(d)] but within a specific time-interval visualizing the short time-period resonant patterns. 
The slow envelope density modulations are evident in panels (a), (d). 
(c) [(f)] Dynamics of the momentum distribution of the density depicted in (a) [(d)],  
displaying the temporal evolution of the instability induced excited modes at $p=0$, $p= \pm k$ and $p= \pm 2k$. 
Here, $\tilde{\psi}(k,t)$ denotes the Fourier transform of $\psi(\theta,t)$}. 
\label{fig:2_omega_2}
\end{figure*}

Importantly, it becomes apparent that by varying the mixing angle $\chi$ and the frequency ratio, it is possible to  reshape the stability diagram: higher-order tongues broaden and become more prominent, while the primary tongue shrinks.  
Particularly, we observe that in the $1{:}2$ case, the broad first Floquet tongue (marked by the red circles) for $\chi=0$ becomes narrower upon gradually increasing $\chi$ and eventually is suppressed, see \cref{fig:mathieu_double}(a)-(e). 
In contrast, the much narrower second Floquet tongue (indicated by the cyan plus marker) is enhanced for larger $\chi$. 
This importantly happens because the latter is the 
primary tongue associated with the resonances of $m=2$
which dominates for mixing angle of $\chi = 90^\circ$.
A similar trend is seen for the $1{:}3$ case, where now  the third Floquet tongue (marked with the red diamond) mainly grows as $\chi$ increases with the first one eliminated and the second predominantly shrinking. Indeed, in this case of $m=3$,
the third tongue becomes the dominant one for $\chi = 90^\circ$.
Concluding, it turns out that the bichromatic driving protocol enables controlled excitation of higher-order resonances, and hence facilitates the tailoring of parametric instabilities, that are otherwise suppressed under a single frequency driving.
This versatility of parametric resonance manipulation within
the frequency-amplitude two-parameter space would be particularly
interesting to realize in forthcoming atomic BEC experiments.

\section{Mean-field pattern formation dynamics} \label{sec:numerical}

\begin{figure*}[htb]
\includegraphics[width=\textwidth]{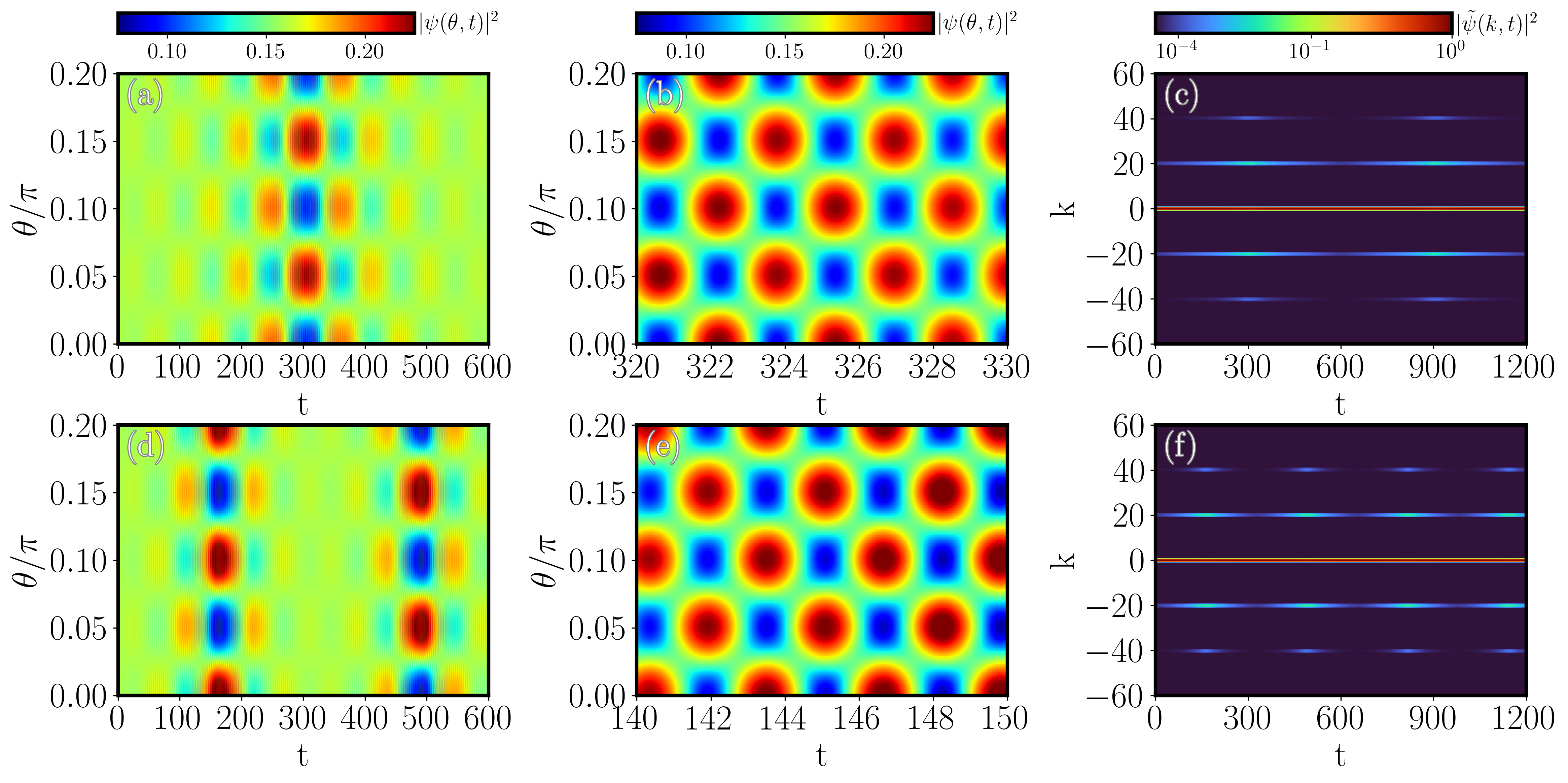} 
\caption{Density evolution of the ring-shaped driven  condensate using the GPE for (a) $\chi = 30^\circ$ and (d) $\chi = 60^\circ$ showing the nucleated nonlinear waves. 
The bichromatic interaction modulation has frequency $\omega_D = 1.34$, amplitude $\alpha = 0.09$, and wavenumber 
$k = 20$. (b) [(e)] Spatio-temporal modulation of the  density patterns shown in (a) [(d)]  within a short time interval depicting the resonance waves. 
(c) [(f)] Time-evolution of the momentum distribution of the density of panel (a) [(f)] showing the participating excited modes at $p=0$, $p= \pm k$ and $p= \pm 2k$. 
Here, $\tilde{\psi}(k,t)$ denotes the Fourier transform of $\psi(\theta,t)$}. 
\label{fig:3_omega_2}
\end{figure*}
\begin{figure}[htb] 
\includegraphics[width=\linewidth]{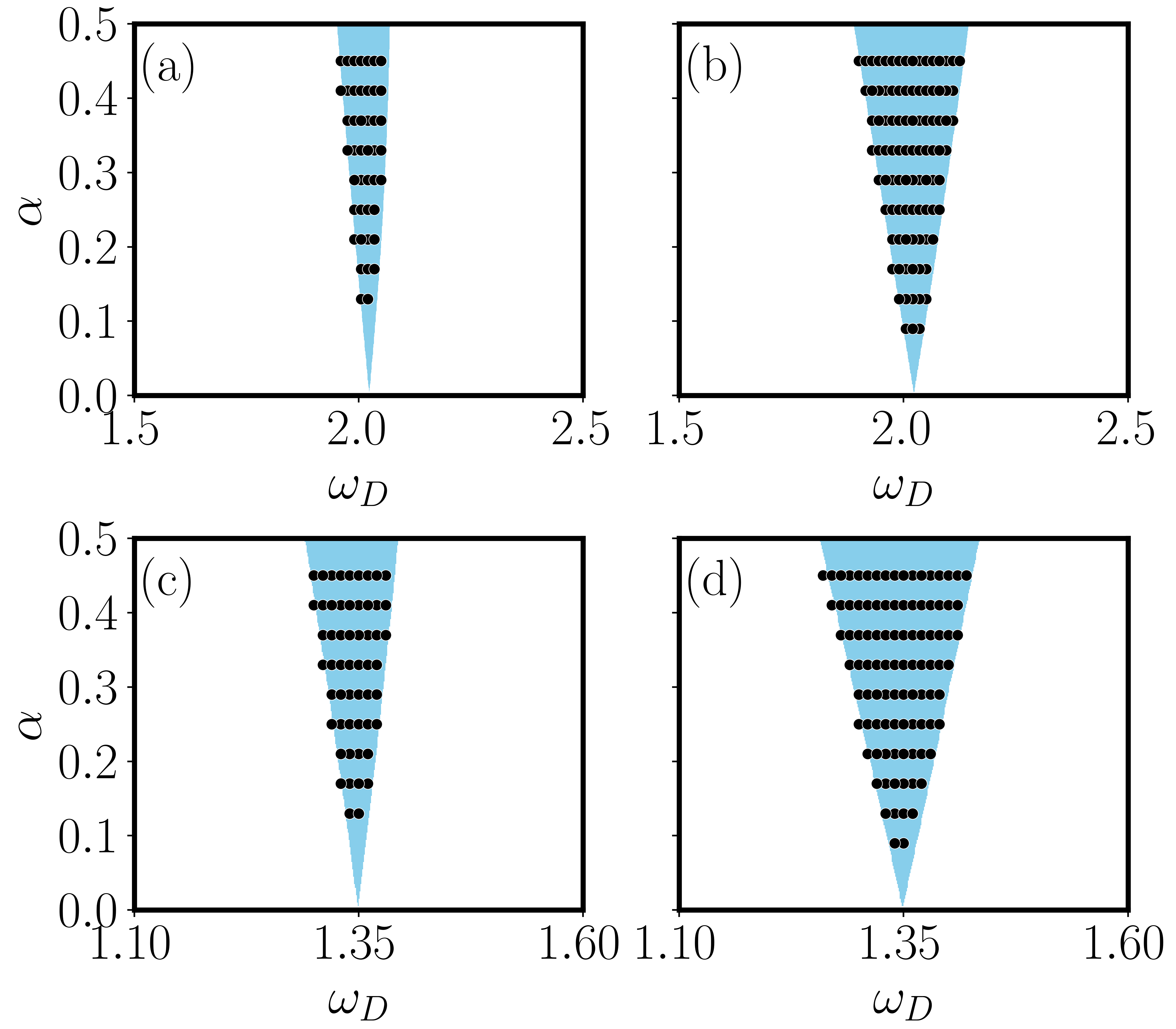} 
\caption{Comparison between the  instability tongues obtained from Floquet theory in the Mathieu equation [Eq.~(\ref{eq:mathieu_equation})] represented by the shaded light blue areas and the GPE simulations (black dots). 
We present the second ($\omega_D=2.02 \approx \Lambda(k)$) and third ($\omega_D=1.35 \approx 2\Lambda(k)/3$) tongues shown also in Fig.~\ref{fig:mathieu_double}. 
The dynamics is induced by bichromatic interaction driving with a frequency ratio of (a), (b) $1{:}2$ and (c), (d) $1{:}3$. 
Different mixing angles are considered corresponding to (a), (c) $\chi=30^\circ$ and (b), (d) $\chi=60^\circ$. 
Excellent agreement between the predictions of the two methods is seen, while relatively small deviations occur for increasing modulation amplitude.}
\label{fig:stability_gpe_mathieu}
\end{figure}

To examine the condensate’s dynamical response to bichromatic driving, we numerically solve the 1D GPE given by Eq.~(\ref{eq:gpe}). 
Particularly, the interaction strength is periodically modulated with two distinct frequency components at a fixed ratio, while the relative weight of each frequency component is regulated by the mixing angle $\chi$. 
The system is initiated into its uniform density 
ground state dubbed as $|\psi_{0}|^2$. Additionally, a small amplitude perturbation of the form $\psi=\psi_{0}[1+\epsilon\mathrm{cos}(k\theta)]$,  where typically $\epsilon=0.01$, is included in our simulations on top the ground state wavefunction to seed faster the ensuing parametric instabilities due to interaction driving.
We remark that such small amplitude perturbation, involving different wave vectors, is commonly contained in a corresponding experiment due to imperfections emanating from noise or thermal fraction~\cite{kwon2021spontaneous}. Hence, in the course of the evolution the ensuing unstable wave vectors will be activated accelerating the emergence of the instability. 
However, the manifestation of the latter  is independent of the inclusion of this perturbation whose ultimate effect is to accelerate the instability growth.

For completeness, we note that the real time evolution of the condensate is computed using the split-step pseudo-spectral method~\cite{Rawat2025, Parker2016,Agrawal2007,yang2010} with a time step $\Delta t=10^{-5}$ and a spatial grid size $\mathcal{M}_{grid}=1024$.

All simulations are carried out over an extended evolution period to ensure that any potential asymptotic behavior of the system is captured.
As expected, if (at least one of) the modulation frequencies lie within the parametrically unstable region determined by the previously described Mathieu tongues [Fig.~\ref{fig:mathieu_double}], dynamical instability takes place. 
This instability is triggered in our case by the time-periodic modulation of the interaction strength. 
It enforces the deformation of the original nearly uniform condensate state into spatially modulated density patterns. 
These modulations correspond to discrete integer wavenumbers, reflecting the periodicity of the ring geometry. 

Figures~(\ref{fig:2_omega_1})-(\ref{fig:3_omega_2}) visualize this unstable behavior on the density level  for frequency ratios $1{:}2$ and $1{:}3$, while the modulation amplitude is fixed to $\alpha = 0.09$. 
More concretely, we first monitor $|\psi(0,t)|^2$ to infer the presence of instability, its strength and time-period.  
This measure is depicted in Fig.~\ref{fig:2_omega_1} (a), (b) for mixing angles $\chi = 30^\circ$ and $\chi = 60^\circ$ respectively while the frequency ratio is held fixed to $1{:}2$ with $\omega_D = 2.03$. {This frequency corresponds to the second instability tongue illustrated in Fig.~\ref{fig:mathieu_double}.} It becomes evident that in both cases, the condensate exhibits rapid temporal density oscillations superimposed with slower envelope modulations, yielding patterns whose amplitude grows and decays periodically in  time. 
The initial exponential growth is the result of the parametric resonance, while the oscillatory behavior is the byproduct of the fully nonlinear dynamics that saturates the instability and 
leads to this longer-term periodic response.

Furthermore, increasing the mixing angle $\chi$ e.g., from $30^\circ$ to $60^\circ$ leads to amplitude enhancement of the density oscillations, indicating that larger 
values of $\chi$ in general yield stronger parametric excitation. 
To infer the temporal oscillation frequency of the patterns we next calculate the respective Fourier spectra of $|\psi(0,t)|^2$ [Fig.~\ref{fig:2_omega_1}(c), (d)].
Besides the trivial peak at $\nu=0$, both spectra feature dominant side peaks at frequencies $ \nu \approx  \pm \Lambda (k)=\pm\omega_D$. These signal the occurrence of the parametric resonance emanating from the second Mathieu tongue, see also Fig.~\ref{fig:mathieu_double}, and the fact that $\chi$ does not alter the temporal frequency of the patterns. 
Additionally, the peak amplitudes in panel (d) are somewhat higher than those in panel (c), providing further confirmation of the increased instability strength for larger $\chi$. However, when plotted on a logarithmic scale,  both peak heights appear nearly identical. This is in full agreement with the Floquet analysis predictions where the instability strength for different $\chi$ has been extracted by calculating the Floquet multipliers of the corresponding generalized Mathieu equation in order to determine the growth rate of the unstable modes (not shown for brevity).

The manifestation of nonlinear dynamical patterns arising in the density evolution of the condensate experiencing bichromatic interaction driving with $1{:}2$ frequency ratio and at different mixing angles is displayed in Fig.~\ref{fig:2_omega_2}(a)-(f) for the unstable wavenumber $k=20$. 
In both cases, the structural deformation of the initial nearly homogeneous background caused by the exponential growth of the parametric  instability induced by the driving can be discerned [Fig.~\ref{fig:2_omega_2}(a), (d)]. 
The periodic re-appearance of an array of spatiotemporal patterns -- resonance waves -- is observed. 
This is related to slow envelope modulations captured by $|\psi(0,t)|^2$, see also Fig.~\ref{fig:2_omega_1}. 
Moreover, each of the aforementioned arrays of patterns, i.e.,
the resonance-induced waves refers to a spatially periodic arrangement of density humps and troughs which alternate within a short time-period as shown in the density evolution of Fig.~\ref{fig:2_omega_2}(b), (e) within first the temporal intervals of their appearance.   
The spatial periodicity of these resonant waves can be inferred by relying on the momentum distribution of the density illustrated for both driving scenaria in Fig.~\ref{fig:2_omega_2}(c), (f). 
It can be seen, that $\pm k=20$ momenta are highly populated and exhibit periodic revivals in the course of the evolution. 
These modes mark the resonant waves spatial periodicity and exemplify that the instability amplifies selectively discrete integer wavenumbers consistent with the ring geometry.  
Additionally, the period of the revivals depends strongly on the mixing angle and it is an imprint of the slow envelope modulations captured by $|\psi(0,t)|^2$.

Turning to the $1{:}3$ bichromatic driving case, we observe a  similar {density response to that described above} with the growth of instability and appearance of the patterns;
see Fig.~\ref{fig:3_omega_2}(a), (d). 
Here, the considered primary driving frequency is $\omega_D = 1.34$ 
which is related to the third Mathieu tongue connected with $\omega_D$ 
in Fig.~\ref{fig:mathieu_double}, but also the first tongue associated
with $3 \omega_D$, since we are involving $m=3$.
Once again, rapid density oscillations of spatially periodic arrays of patterns, are clearly visualized in Fig.~\ref{fig:3_omega_2}(b), (e), superimposed with a slow periodic amplitude modulation. 
As expected, this dynamical behavior is consistent with the evolution of $|\psi(0,t)|^2$ demonstrated in Fig.~\ref{fig:2_omega_1}(e), (f). 
Inspecting $|\psi(0,t)|^2$, it becomes clear that the amplitude and temporal frequency of the periodic density modulation depend on the mixing angle. 
Namely, the oscillation amplitude is enhanced for larger $\chi$, indicating increasing strength of the parametric excitation, {while the corresponding oscillation period decreases in accordance with
Eq.~\eqref{eq:omega_n} for $m \omega_D$ with $m=3$.}
Relying on the spectrum of $|\psi(0,t)|^2$, see Fig.~\ref{fig:2_omega_1}(g), (h), we verify the dominant nature
of the population associated with the frequency $\nu= \pm \Lambda (k) = 1.5 \omega_D$ independently of $\chi$. 

On the other hand, a momentum space analysis of the density shown in Fig.~\ref{fig:3_omega_2}(c), (f) reveals the selective amplification of the $k = \pm 20$ modes which is in line with the used unstable wavenumber inducing pattern formation.  
Furthermore, as in the $1{:}2$ case, increasing the mixing angle $\chi$ enhances the instability growth with the patterns arising at smaller evolution times.

Also, the temporal period of the excited modes decreases for larger $\chi$, see \cref{fig:3_omega_2}(c), (f).

To further corroborate the validity of the instability phase diagrams across the $\alpha$-$\omega_D$ planes, we subsequently compare the analytically predicted resonance tongues extracted from the Mathieu equation [Eq.~(\ref{eq:mathieu_equation})] with those predicted by dynamical GPE simulations. 
In the latter, we identify the emergence of unstable dynamics when the condensate density grows rapidly (effectively exponentially) -- typically at least exceeding ~$10 \%$ of its initial background value -- due to the  presence of a parametric resonance.   
Such a criterion has also been used in Ref.~\cite{Turmanov2020} to quantify the generation of density waves.

Figure~\ref{fig:stability_gpe_mathieu} summarizes the above described comparison for selected frequency ratios and mixing angles. 
The scenario with driving frequency ratio $1{:}2$ 
and $\chi=30^\circ$ ($\chi=60^\circ$) associated with the second instability tongue 
(for $\omega_D$ and first one for $2 \omega_D$) shown in  Fig.~\ref{fig:mathieu_double}(b) [(d)] is depicted in Fig.~\ref{fig:stability_gpe_mathieu}(a) [(b)].  
Similarly, the scenario following a $1{:}3$ frequency ratio at $\chi=30^\circ$ [$\chi=60^\circ$] related to the third  tongue of  Fig.~\ref{fig:mathieu_double}(g) [(i)] is presented in Fig.~\ref{fig:stability_gpe_mathieu}(c) [(d)]; recall that
these are the first tongues, respectively, for the cases of $m=2$ and $m=3$. 
In all cases, shaded light-blue regions determine the boundaries of the instability tongues predicted by Floquet analysis of the Mathieu equation [Eq.~(\ref{eq:mathieu_equation})], and black dots designate the presence of instability captured by the GPE simulations.

Overall, excellent agreement between the analytical predictions and the numerical results can be readily deduced especially at relatively small driving amplitudes. 
This confirms that the generalized Mathieu equation provides a robust and reliable framework for delineating the instability boundaries in such driven condensates over a wide range of parameters. Such observations have been reported in previous works concentrated to single frequency driving~\cite{Staliunas2002,maity2020parametrically,kwon2021spontaneous}. 
Moreover, it can be seen that as the mixing angle $\chi$ increases from $30^\circ$ to $60^\circ$, both the second tongue in the $1{:}2$ case [Fig.~\ref{fig:stability_gpe_mathieu}(a), (b)] and the third tongue in the $1{:}3$ one  [Fig.~\ref{fig:stability_gpe_mathieu}(c), (d)] broadens significantly, while the primary (widest) tongue becomes narrower (not shown). 
This behavior is consistent with the trend shown in  \cref{fig:mathieu_double}. 
Finally, small deviations in the instability boundaries as captured by the two methods arise for larger driving amplitudes, with the GPE predicting slightly wider boundaries. This can be traced back to the validity region of the generalized Mathieu equation which is obtained through linearization of {the GPE} and is, thus, expected to be progressively less accurate for increasingly large amplitude; see also Sec.~\ref{sec:mathieu}.

\section{Five Mode Approximation}\label{sec:five_Mode}
\begin{figure*}[!htb] 
\includegraphics[width=\linewidth]{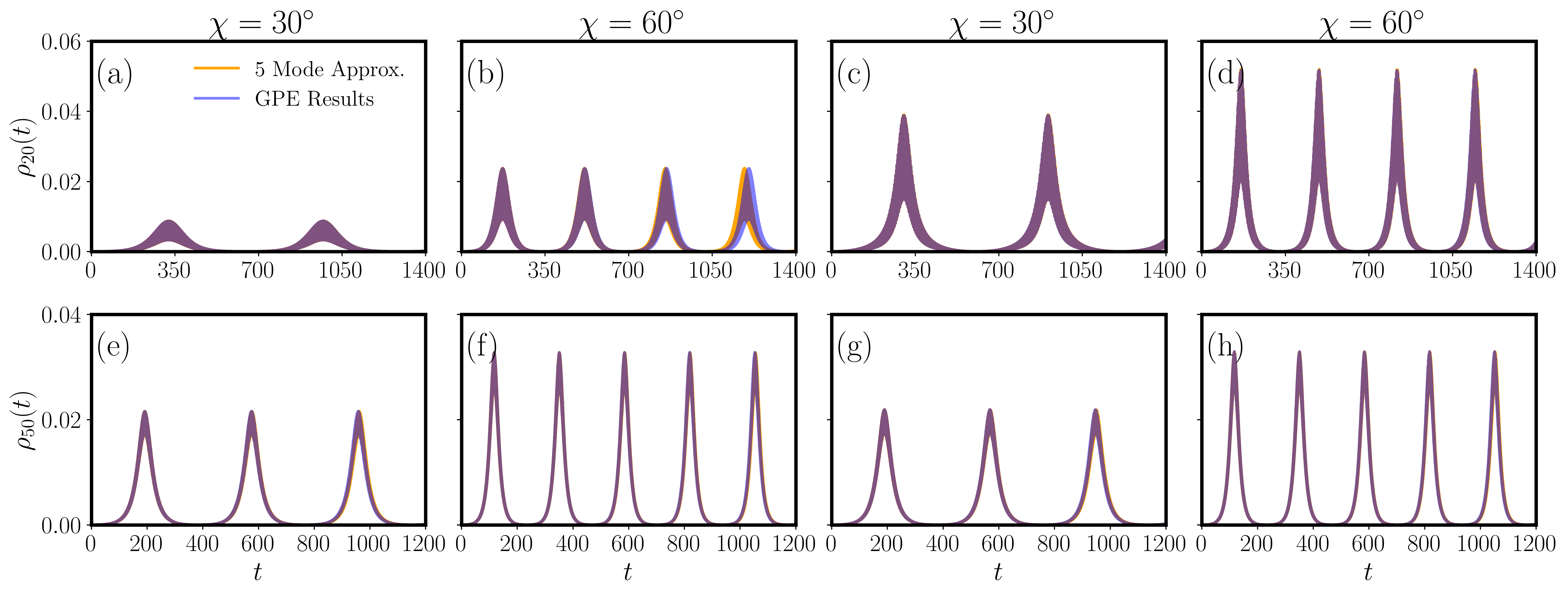} 
\caption{Dynamics of $\rho_k(t)$ for frequency ratios (a), (b), (e), (f) $1{:}2$ and (c), (d), (g), (h) $1{:}3$ at mixing angles (a), (c), (e), (g) $\chi=30^\circ$ and (b), (d), (f), (h) $\chi=60^\circ$. 
Results from both the GPE [Eq.~(\ref{eq:gpe})] and the five-mode approximation [Eq.~(\ref{eq:5mode_hamilt})] are shown (see legends). 
As it can be seen, the five-mode approximation accurately follows the GPE results in the course of the evolution for both higher-modes (lower panels) and lower unstable wavenumbers (upper panels). 
The primary driving frequency corresponds to (a), (b) $\omega_D=2.03 \approx \Lambda(k=20)$, (c), (d) $\omega_D=1.34 \approx 2\Lambda(k=20)/3$, (e), (f) $\omega_D=8.79  \approx \Lambda(k=50)$ and (g), (h) $\omega_D=5.86 \approx 2\Lambda(k=50)/3$. 
The considered unstable wavenumber is $k=20$ (upper panels) and $k=50$ (lower   panels). 
In all cases, the used driving amplitude is $\alpha=0.09$.}
\label{fig:3mode_k50_amp}
\end{figure*}

Next, in order to capture the essential features of the observed nonlinear dynamics underlying bichromatically driven parametric instabilities, we  derive a reduced five-mode model. This is motivated by the GPE simulations presented in Sec.~\ref{sec:numerical}, see also Fig.~\ref{fig:2_omega_2} (e), (f) and Fig.~\ref{fig:3_omega_2} 
(e), (f), where the condensate evolution is dominated by the {$p=0$ mode together with the resonantly unstable side modes at $p=\pm k$. Retaining only these modes provides a reduced effective description of the condensate response}, argued to be sufficient for the description of the effects caused by  monochromatic driving~\cite{zhu2019}. 
In our case, however, it turns out that the emergent dynamics induced by bichromatic interaction driving requires the involvement of five modes. 
Indeed, we find that while a corresponding three-mode model is capable of accurately reproducing the dynamics for relatively high-wavenumber excitations (e.g., $k=50$), its predictions deviate significantly from the GPE results 
at lower wavenumbers such as  $k=20$.

Therefore, to construct a model with a broader range of applicability
that is able to adequately describe the impact of the ensuing parametric instability over the entire range of participating  wavenumbers, including the low-$k$ regime, we consider a five-mode truncation of the underlying GPE dynamics. 
As such, this truncation scheme contains the modes $\eta = 0, \pm k, \pm 2k$. 
Each of these modes is represented by a complex amplitude, $\psi_\eta$, defined through the Fourier expansion of the condensate wavefunction given by  
$\langle \hat{\psi} \rangle = (1/\sqrt{2\pi})\sum_\eta \psi_\eta(t) e^{i\eta \theta}$.  
Here, $\psi_\eta(t) $ denotes the amplitude of the mode characterized by momentum $\eta$ which takes values within the momentum interval $\{0, \pm k, \pm 2k\}$. 
The occupation number of the $\eta$-mode refers to $n_\eta = |\psi_\eta|^2$. 
Accordingly, the truncated five-mode Hamiltonian is expressed as 
\begin{widetext}
\begin{align}
H = & \; \lambda k^2 \Big[n_k + n_{-k} + 4(n_{2k} + n_{-2k})\Big] +
\frac{g(t)}{2N} \Big[ (n_0^2 + n_k^2 + n_{-k}^2 + n_{2k}^2 + n_{-2k}^2) + 
4\big(n_0 n_k + n_0 n_{-k} + n_0 n_{2k} \notag \\ 
& + n_0 n_{-2k} + n_k n_{-k} + n_k n_{2k} + n_k n_{-2k} + n_{-k} n_{2k} + n_{-k} n_{-2k} + n_{2k} n_{-2k} \big) +
2\big( \psi_{-2k}^* \psi_0^* \psi_{-k} \psi_{-k}  \notag \\ 
& + \psi_{-k}^* \psi_{-k}^* \psi_0 \psi_{-2k} \big) +  4\big( \psi_{-2k}^* \psi_k^* \psi_{-k} \psi_0 +
\psi_0^* \psi_{-k}^* \psi_k \psi_{-2k} \big)  + 4\big( \psi_{-2k}^* \psi_{2k}^* \psi_{-k} \psi_k + 
\psi_k^* \psi_{-k}^* \psi_{2k} \psi_{-2k} \big) \notag \\ 
&+ 2\big( \psi_{-2k}^* \psi_{2k}^* \psi_0 \psi_0 + 
\psi_0^* \psi_0^* \psi_{2k} \psi_{-2k} \big)   + 2\big( \psi_{-k}^* \psi_k^* \psi_0 \psi_0 + \psi_0^* \psi_0^* \psi_k \psi_{-k} \big) + 
4\big( \psi_{-k}^* \psi_{2k}^* \psi_0 \psi_k  \notag \\ 
& + \psi_k^* \psi_0^* \psi_{2k} \psi_{-k} \big)   +
2\big( \psi_0^* \psi_{2k}^* \psi_k \psi_k + \psi_k^* \psi_k^* \psi_{2k} \psi_0 \big) \Big],
\label{eq:5mode_hamilt}
\end{align}
\end{widetext}

\begin{figure*}[!htb]
\includegraphics[width=\textwidth]{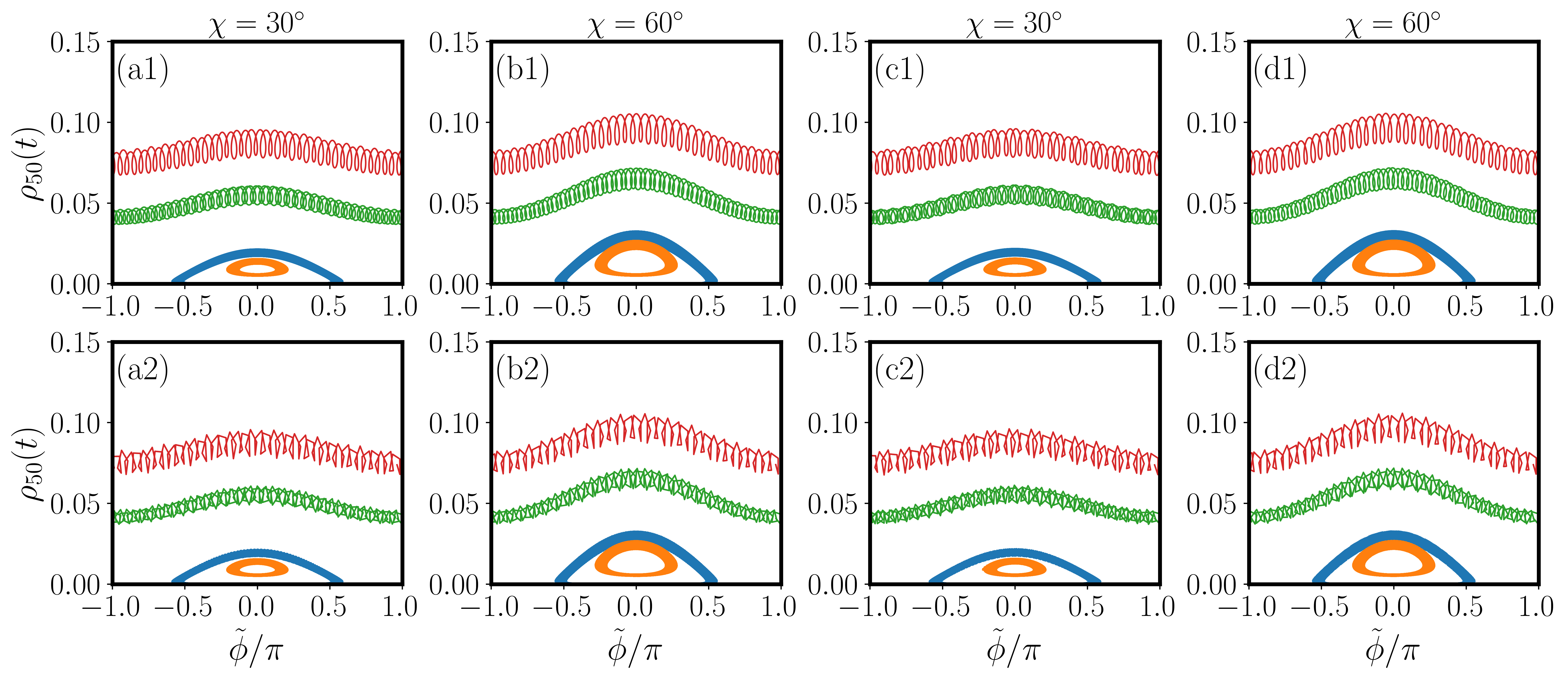} 
\caption{Phase-plane portraits of the $\rho_{50}$ mode  dynamics for a driven condensate under bichromatic interaction driving  extracted using (a1)-(d1) the five-mode  approximation and (a2)-(d2) the GPE simulations. 
Closed orbits indicate the presence of parametric resonances. Good agreement is observed between the phase-plane contours of the GPE and that from the five-mode model. 
The driving characteristics refer to (a1), (b1), (a2), (b2) $1{:}2$ frequency ratio with amplitude $\alpha = 0.09$, primary frequency $\omega_D = 2.03\approx \Lambda(k)$, 
and (c1), (d1), (c2), (d2) $1{:}3$ ratio, amplitude $\alpha = 0.09$, and primary frequency $\omega_D = 1.34\approx 2\Lambda(k)/3$. The mixing angles are shown in the legends and the unstable wavenumber is $k = 50$. 
For the $1{:}2$ ($1{:}3$) frequency ratio it holds that $\tilde{\phi} = \phi_k - 2\omega_D t$ ($\tilde{\phi} = \phi_k - 3\omega_D t$). 
}
\label{fig:3mode_phase_k50}
\end{figure*}

Considering $\psi_{\eta} = \sqrt{n_{\eta}(t)} e^{i \theta_{\eta}(t)}$, and using the complex Hamiltonian equation of motion, it is possible to derive the corresponding equations of motion for the classical fields $\{ \rho_{k}, \rho_{2k}, \phi_{k}, \phi_{2k} \}$ satisfying 
\begin{subequations}
\begin{align}
\dot{\rho}_k & =  2g(t) \rho_k \big[ \rho_0 \sin(\phi_k) + 2\rho_{2k} \sin (\phi_{k} - \phi_{2k}) \notag \\
              &  + 2\sqrt{\rho_0 \rho_{2k}} \sin (\phi_k - \phi_{2k}/2) \big], 
\label{eq:rhok} \\ \nonumber \\
\dot{\rho}_{2k} & = 2g(t) \big[ \rho_0 \rho_{2k} \sin(\phi_{2k}) + 2\rho_k \sqrt{\rho_0 \rho_{2k}} \sin(\phi_{2k}/2) \notag \\ 
& + 2\rho_{2k} \rho_k \sin(\phi_{2k} - \phi_{k}) + \rho_k \sqrt{\rho_0 \rho_{2k}} \sin (\phi_{2k}/2 - \phi_{k}) \big], 
\label{eq:rho2k} \\
\dot{\phi}_k &= 2\lambda k^2 + 2g(t)(\rho_0 - \rho_k)  + 2g(t) \Big[ \big(\rho_0 - 2\rho_k\big)\cos\phi_k \notag \\
& + 2\rho_{2k}\big(\cos(\phi_{2k} - \phi_k) - \cos\phi_{2k}\big)  \notag \\
&+ 4(\rho_0 - \rho_k)\sqrt{\frac{\rho_{2k}}{\rho_0}}\cos\left({\phi_{2k}/2}\right) \notag \\
& + 2(\rho_0 - \rho_k)\sqrt{\frac{\rho_{2k}}{\rho_0}}\cos\left(\phi_k - {\phi_{2k}/2}\right) \Big], 
\label{eq:phik} \\
\dot{\phi}_{2k} &=  8\lambda k^2 + 2g(t)(\rho_0 - \rho_{2k}) + 2g(t) \Big[ (\rho_0 - 2\rho_{2k})\cos\phi_{2k}  \notag \\
& + 2\rho_k\big( \cos(\phi_{2k}-\phi_k)-  \cos\phi_k\big)  + 2A\rho_k\cos\left({\phi_{2k}/2}\right) \notag \\
& + A\rho_k\cos\left(\phi_k - {\phi_{2k}/2}\right) \Big]. 
\label{eq:phi2k}
\end{align}
\end{subequations}
In these expressions, $A \equiv \sqrt{\rho_{0}/\rho_{2k}} - 2\sqrt{\rho_{2k}/\rho_{0}}$, while $\rho_{0}=n_{0}/N$, $\rho_{k}=n_{k}/N$, $\rho_{2k}=n_{2k}/N$ (with $N=n_0+n_{k}+n_{-k}+n_{2k}+n_{-2k}$) and $ \phi_{k} = 2\theta_{0} - \theta_{k}-\theta_{-k}$, $\phi_{2k} = 2\theta_{0} - \theta_{2k}-\theta_{-2k}$.
Also, we have assumed that $n_k=n_{-k}$, $n_{2k}=n_{-2k}$, $\theta_{k}=\theta_{-k}$, $\theta_{2k}=\theta_{-2k}$.

To assess the validity of this five-mode model and gain further analytical insights, we subsequently compare its predictions to 
the ones obtained from the full GPE simulations. 
Specifically, we compute the mode populations $\rho_k = |\tilde{\psi}(k, t)|^2$, where, $\tilde{\psi}(k, t)$ represents the Fourier transform of  $\psi(\theta, t)$,  
directly from the GPE,
and from the reduced five-mode model described by Eqs.~(\ref{eq:rhok})-(\ref{eq:phi2k}). 
The time-evolution of representative high-mode ($k=50$) populations is presented in \Cref{fig:3mode_k50_amp} in the cases of $1{:}2$ [Fig.~\ref{fig:3mode_k50_amp}(e), (f)] and $1{:}3$ [Fig.~\ref{fig:3mode_k50_amp}(g), (h)] frequency ratios for different mixing angles. 
It is evident that the five-mode model is able to adequately capture the shape, amplitude and two-frequency oscillation trend of $\rho_{50} $ independently of the driving frequency ratio
when compared to the GPE results.
We remark that a similar agreement can be inferred by utilizing a corresponding three-mode model (not shown) for the case of $k=50$ (but not of that of $k=20$)
although in this case the amplitude of $\rho_{50} $ is slightly overestimated relatively to the GPE predictions. 
Such deviations that become significant for lower population modes examined below are traced back to the omission of higher-order mode couplings within the respective three-mode reduced model;
in the GPE some
of the conserved total mass is transferred to higher momenta. 

The dynamics of lower population modes, e.g., $\rho_{20}$, within the five mode approximation is illustrated in \cref{fig:3mode_k50_amp}(a)-(d) for different driving frequency ratios and mixing angles. 
It can be readily seen, that both the oscillation amplitude and frequencies of $\rho_{20}$ are in close agreement to each other in both the reduced and GPE approaches irrespectively of the driving frequency ratio and mixing angles. 

The above-discussed agreement clearly highlights the ability of the simplified five-mode model to capture key features of the parametric resonance dynamics. 
It thus provides a reliable and computationally efficient framework,
where appropriate, for describing the onset and nonlinear evolution of parametric instabilities in bichromatically driven condensates. Certainly, the interplay of higher-mode couplings and multifrequency driving protocols is an interesting topic deserving further study in the future.

To further characterize the system's nonlinear dynamics we finally extract the underlying phase-plane contours using both the five mode [\cref{fig:3mode_phase_k50}(a1)-(d1)] and the GPE  [\cref{fig:3mode_phase_k50}(a2)-(d2)] methods. 
We define \(\tilde{\phi} = \phi -2\omega_D t\) for the $1{:}2$ frequency ratio and \(\tilde{\phi} = \phi -3\omega_D t\) for the  $1{:}3$ ratio and characterize the long-term dynamics of the system in the \(\rho_k - \tilde{\phi}\) plane for various mixing angles. 
In all cases, the phase space dynamics contains closed orbits and open trajectories. 
The self-trapped closed contours are characteristic of the presence of the parametric resonances~\cite{zhu2019}. 
They reflect, on the one hand, larger density variations,
as we observed in the growth dynamics, but also indicate
the practically oscillatory nature of the corresponding
dynamics which is mirrored in the oscillations that we
also noticed in the GPE simulations.
Accordingly, the phase portraits from the reduced five-mode model are in excellent  agreement with the GPE ones, confirming once more that the essential features of the phase-space dynamics are well captured by this simplified description.
In order to obtain these phase portraits at the GPE level,
we have projected the full field dynamics to the 
$(\rho_{50},\tilde{\phi}_{50})$ plane, using the
Fourier decomposition of the GPE field. 
We also find a good agreement in the phase-space dynamics occurring for lower modes (not shown for brevity) similarly to what
we observed in the $\rho_{k}$ dynamics above in Fig.~\ref{fig:3mode_phase_k50}.

\section{Conclusions and Future Perspectives} \label{sec:con}

We have investigated the emergence of  pattern formation
as a result of parametric resonance in a one-dimensional Bose–Einstein condensate confined to a ring and subjected to bichromatic modulation of the interaction strength. 
Utilizing linear stability analysis of a homogeneous density
state, we have analytically extracted a suitable 
generalized Mathieu equation. The use of  Floquet theory 
then allows to map out the stability phase diagram with respect to characteristics of the driving protocol, namely the driving frequency, amplitude, and the mixing angle defined herein,
based on analogies to the fluid mechanics literature. 

It is found that the bichromatic driving gives rise to different resonance tongues that correspond to parametric instabilities. 
Moreover, the ratio of the two driving frequencies enables to selectively broaden higher-order tongues while narrowing the primary one. 
This scheme facilitates the control of the excited instability modes of the condensate. Accordingly, we developed a 
homotopic protocol that possesses as special case limits
the monochromatic drive cases (at two different frequencies)
as studied before, but also expands upon them enabling a
bichromatic drive that has been eminently shown to be
experimentally accessible. The analysis provided shows how
the linearized stability landscape can be sculpted on the
basis of the parametric freedom within such a drive.

Our linear stability results are corroborated by suitable time-dependent Gross–Pitaevskii simulations whose predictions, e.g., in terms of the location of the resonance tongues and the impact of the mixing angle on the instabiity strength are in 
good agreement, where appropriate, 
with the aforementioned analytical approach.  
Additionally, they reveal the development of density modulations  characterized by integer wavenumber
consistent with the presence of the ring geometry. 
To gain deeper insights into the nonlinear stage of the unstable dynamics, we constructed also a reduced {five-mode} model which is able to capture the main instability induced nonlinear features. 
The latter refer, for instance, to the onset of the instability, 
and, importantly, the nonlinearity-induced saturation and the
resulting  oscillations of the mode populations.
The phase-space structure obtained from this {five-mode} model
was found to be in very good agreement with full mean-field  simulations, again in the regime where the {five-mode} approximation
is supported by the partial differential equation dynamics.

These findings highlight the potential of multi-frequency external modulations as a powerful tool for engineering and controlling pattern formation and instabilities in quantum fluids. 
Among the most immediate next steps, one can note 
the consideration of different frequency combinations,
the incorporation of dissipation (e.g., from three-body
losses) and observation of how the instability tongues
are affected therefrom. Moreover, the potential extensions
(to more modes) and the detailed study of the mathematical
structure of few-mode models could also provide further 
insights into the system.

An immediate key generalization would be to explore parametric instabilities in  higher dimensions first in single-component and afterwards in spinor settings. This is expected to give access to enriched and far more complex patterns~\cite{wang2025parametric,fujii2024stable,liebster2025observation} which may even possess supersolid-like phase characteristics as in the recent experiment of Ref.~\cite{liebster2025supersolid}. 
Certainly, investigating the impact of parametric instabilities in more exotic phases-of-matter such as droplets~\cite{luo2021new,mistakidis2023few} encapsulating quantum fluctuations and suffering self-evaporation is also desirable. 
Another intriguing pathway would be to examine the correlation properties of such instabilities as was done in the experiment of Ref.~\cite{nguyen2019parametric} using sophisticated many-body computational methods~\cite{Bougas_driving}. 

\acknowledgments
We gratefully acknowledge our Param-Ishan and Param Kamrupa supercomputing facility (IITG), where all numerical simulations were performed. P.M. gratefully acknowledges  research fellowship from MoE, Government of India. S.I.M. acknowledges
financial support from the Startup Fund of the 
Curators of the University of Missouri of Science
and Technology, in connection with the 
Department of Physics.
This research was supported by the U.S. National Science Foundation under the awards DMS-2204702 and PHY-2408988 (PGK). This research was partly conducted while P.G.K. was 
visiting the Okinawa Institute of Science and
Technology (OIST) through the Theoretical Sciences Visiting Program (TSVP). 
This work was also 
supported by a grant from the Simons Foundation
[SFI-MPS-SFM-00011048, P.G.K].
 P.G.K. is also grateful to Professors
Priya Subramanian and Laurette Tuckerman for relevant discussions.

\appendix

\counterwithin{figure}{section}
\section{Floquet Analysis}
\label{A:1}

To analyze the stability of the parametrically driven condensate, we recall that the perturbation dynamics [Eq.~(\ref{eq:mathieu_equation}) in the main text] can be written as a Mathieu-type equation containing a time-dependent coefficient of fundamental period $T = 2\pi/\omega_D$. It is therefore desirable to employ Floquet theory atop the Mathieu for exploring the system's stability characteristics. 
In particular, the Mathieu equation is a special case of Hill’s equation~\cite{magnus2004hill,eckardt2017colloquium, bukov2015universal}
\begin{equation}
\frac{d^2u}{dt^2} + c(t)u = 0, \quad \text{with} \quad c(t+T) = c(t),
\label{eq:a1}
\end{equation}
where $c(t)$ is a time periodic function. For our system, it holds that $c(t) = \Omega^2(k) + 2 \alpha \lambda k^2 \big[ \cos(\chi)\cos(\omega_D t) +\sin(\chi)\cos(m\omega_D t)\big]$ and $T = 2\pi/\omega_D$.

To recast Eq.~(\ref{eq:a1}) into a system of first-order ordinary differential equations, we employ the new variables $u_1 = u$ and $u_2 = \frac{du}{dt}$, yielding:
\begin{equation}
\frac{d}{dt}
\begin{bmatrix}
u_1 \\
u_2
\end{bmatrix}
=
\begin{bmatrix}
0 & 1 \\
-c(t) & 0
\end{bmatrix}
\begin{bmatrix}
u_1 \\
u_2
\end{bmatrix}.
\end{equation}

To proceed, we define the fundamental matrix solution composed of two linearly independent solution vectors:
\begin{equation}
\begin{bmatrix}
u_{11}(t) \\
u_{12}(t)
\end{bmatrix}
\quad \text{and} \quad
\begin{bmatrix}
u_{21}(t) \\
u_{22}(t)
\end{bmatrix},
\end{equation}
with the initial conditions:
\begin{equation}
\begin{bmatrix}
u_{11}(0) \\
u_{12}(0)
\end{bmatrix}
=
\begin{bmatrix}
1 \\
0
\end{bmatrix},
\quad
\begin{bmatrix}
u_{21}(0) \\
u_{22}(0)
\end{bmatrix}
=
\begin{bmatrix}
0 \\
1
\end{bmatrix}.
\end{equation}

By evaluating this fundamental matrix solution at $t = T$, we obtain the monodromy matrix $M$:
\begin{equation}
F =
\begin{bmatrix}
u_{11}(T) & u_{21}(T) \\
u_{12}(T) & u_{22}(T)
\end{bmatrix}.
\end{equation}

According to Floquet theory, the system's stability depends on the eigenvalues, also known as Floquet multipliers, of the matrix $F$, which satisfy the characteristic equation:
\begin{equation}
	f^2 - (\text{tr}\,F)f + \det F = 0,
\end{equation}
where $\text{tr}\,F$ and $\det F$ denote the trace and determinant of $F$, respectively. For this system, it is known that $\det F = 1$, allowing us to simplify the characteristic equation:
\begin{equation}
f^2 - (\text{tr}\,F)f + 1 = 0
\end{equation}

The solutions to this quadratic equation are given by:
\begin{equation}
f_{1,2} = \frac{\text{tr}\,F \pm \sqrt{(\text{tr}\,F)^2 - 4}}{2}.
\end{equation}

Instability occurs when either eigenvalue has magnitude greater than one, i.e., $\vert f_{1,2}\vert > 1$~\cite{magnus2004hill,ince1956ode}. The boundary between stable and unstable behavior is determined by parameter values for which $\vert\text{tr}\,F\vert = 2$. In particular, when $\text{tr}\,F = 2$, both eigenvalues are $f_{1,2} = 1$.

\bibliography{citation.bib} 
\end{document}